\documentclass[aip,jcp,reprint,noshowkeys,superscriptaddress]{revtex4-1}
\usepackage{graphicx,dcolumn,bm,xcolor,microtype,multirow,amscd,amsmath,amssymb,amsfonts,physics,wrapfig,txfonts,siunitx}
\usepackage[version=4]{mhchem}

\newcommand{\ie}{\textit{i.e.}}

\usepackage[normalem]{ulem}

\newcommand{\SupInf}{\textcolor{blue}{Supporting Information}}

\newcommand{\rev}[1]{\textcolor{black}{#1}}

\newcommand{\fnm}{\footnotemark}
\newcommand{\fnt}{\footnotetext}
\newcommand{\tabc}[1]{\multicolumn{1}{c}{#1}}
\newcommand{\QP}{\textsc{quantum package}}

\newcommand{\hI}{\Hat{1}}
\newcommand{\hT}{\Hat{T}}
\newcommand{\hH}{\Hat{H}}
\newcommand{\bH}{\Bar{H}}
\newcommand{\ERI}[2]{v_{#1}^{#2}}

\newcommand{\EFCI}{E_\text{FCI}}
\newcommand{\ECC}{E_\text{CC}}
\newcommand{\EVCC}{E_\text{VCC}}

\usepackage[
	colorlinks=true,
    citecolor=blue,
    breaklinks=true
	]{hyperref}
\begin{document}

\newcommand{\LCPQ}{Laboratoire de Chimie et Physique Quantiques (UMR 5626), Universit\'e de Toulouse, CNRS, UPS, France}

\title{Excited States From State Specific Orbital Optimized Pair Coupled Cluster}

\author{F\'abris Kossoski}
\email{fkossoski@irsamc.ups-tlse.fr}
\affiliation{\LCPQ}
\author{Antoine Marie}
\affiliation{\LCPQ}
\author{Anthony Scemama}
\affiliation{\LCPQ}
\author{Michel Caffarel}
\affiliation{\LCPQ}
\author{Pierre-Fran\c{c}ois Loos}
\email{loos@irsamc.ups-tlse.fr}
\affiliation{\LCPQ}

% Abstract
\begin{abstract}
The pair coupled cluster doubles (pCCD) method (where the excitation manifold is restricted to electron pairs) has a series of interesting features.
Among others, it provides ground-state energies very close to what is obtained with doubly-occupied configuration interaction (DOCI),
but with polynomial cost (compared with the exponential cost of the latter).
Here, we address whether this similarity holds for excited states, by exploring the symmetric dissociation of the linear \ce{H4} molecule.
When ground-state Hartree-Fock (HF) orbitals are employed, pCCD and DOCI excited-state energies do not match, a feature that is assigned to the poor HF reference.
In contrast, by optimizing the orbitals at the pCCD level (oo-pCCD) specifically for each excited state, the discrepancies between pCCD and DOCI decrease by one or two orders of magnitude.
Therefore, the pCCD and DOCI methodologies still provide comparable energies for excited states, but only if suitable, state-specific orbitals are adopted.
We also assessed whether a pCCD approach could be used to directly target doubly-excited states, without having to resort to the equation-of-motion (EOM) formalism.
In our $\Delta$oo-pCCD model, excitation energies were extracted from the energy difference between separate oo-pCCD calculations for the ground state and the targeted excited state.
For a set comprising the doubly-excited states of \ce{CH+}, \ce{BH}, nitroxyl, nitrosomethane, and formaldehyde, we found that $\Delta$oo-pCCD provides quite accurate excitation energies,
with root mean square deviations (with respect to full configuration interaction results) lower than CC3 and comparable to EOM-CCSDT, two methods with much higher computational cost.
\bigskip
\begin{center}
        \boxed{\includegraphics[width=0.4\linewidth]{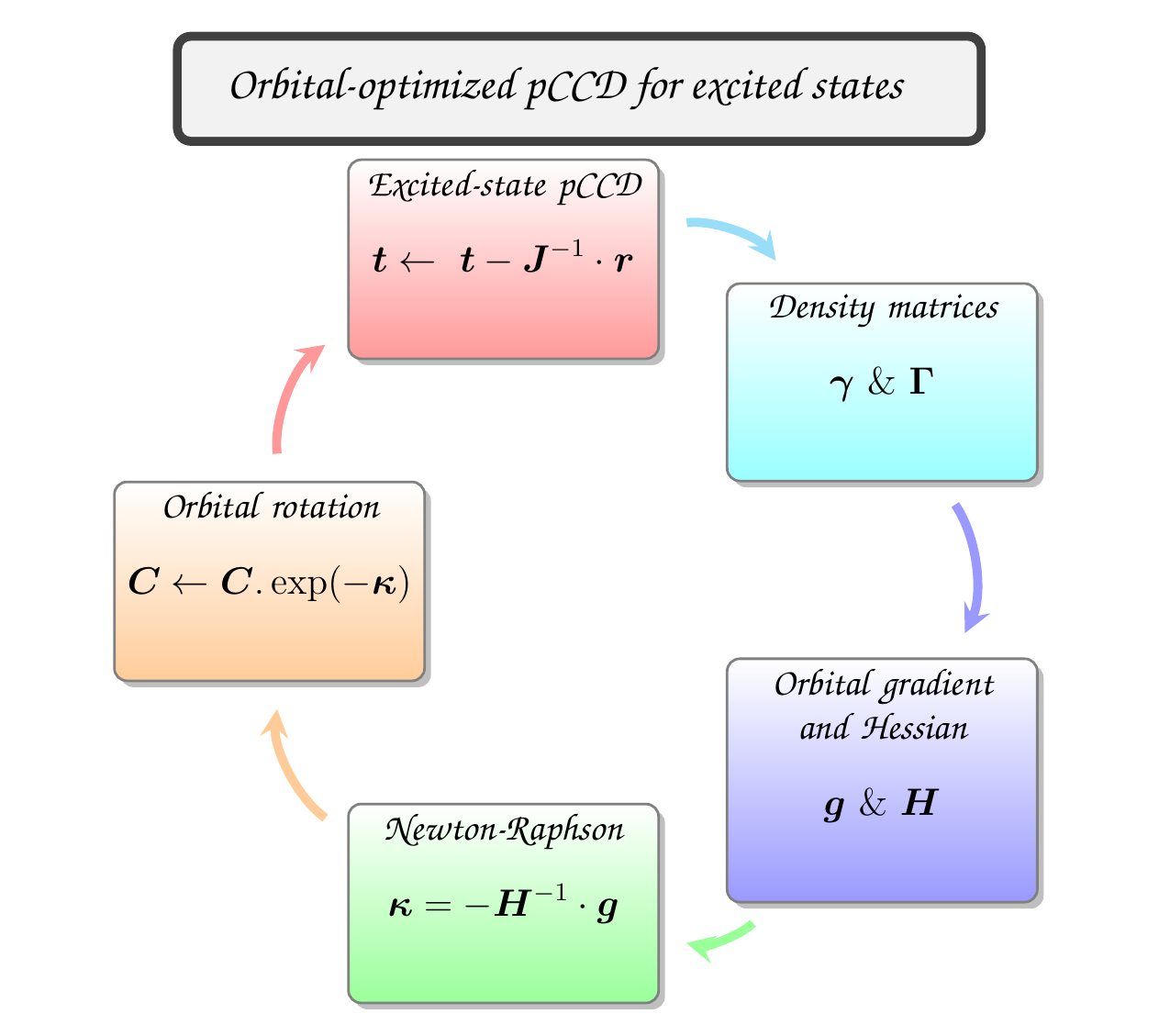}}
\end{center}
\bigskip
\end{abstract}

% Title
\maketitle

%%%%%%%%%%%%%%%%%%%%%%%%%%%%%%%%%%%%%%%%%%%%%%%%%%
\section{Coupled cluster for ground states}
\label{sec:GSCC}
%%%%%%%%%%%%%%%%%%%%%%%%%%%%%%%%%%%%%%%%%%%%%%%%%%

\rev{The coupled cluster (CC) family of methods \cite{Cizek_1966,Paldus_1972,Crawford_2000,Bartlett_2007,Shavitt_2009} is one of the most successful wave function approaches for the description of chemical systems. \cite{Pople_1978,Bartlett_1978,Purvis_1982,Scuseria_1987a,Scuseria_1988,Scuseria_1989}
In particular, low-order truncated CC methods, such as CC with singles, doubles and perturbative triples CCSD(T), \cite{Purvis_1982,Raghavachari_1989} properly describe weak correlation, while inclusion of higher-order excitations is required for strongly correlated systems.}

In CC theory, the exponential excitation operator 
\begin{equation}
	e^{\hT} = \hI + \hT + \frac{\hT^2}{2!} + \frac{\hT^3}{3!} + \ldots,
\end{equation}
with 
\begin{equation}
	\hT = \sum_{n=1}^N \hT_n,
\end{equation}
(where $N$ is the number of electrons) acts on a (normalized) single Slater determinant $\ket{\Phi}$ [such as Hartree-Fock (HF)] to convert it into the exact wave function 
\begin{equation}
	\ket{\Psi} = e^{\hT} \ket{\Phi}.
\end{equation}
The $n$th excitation operator $\hT_n$ is defined, in second-quantized form, as
\begin{equation}
	\label{eq:T_k}
	\hT_n = \frac{1}{(n!)^2} \sum_{ij\dots} \sum_{ab\dots} t_{ij\dots}^{ab\dots} c_a^{\dag} c_b^{\dag} \dots c_j c_i,
\end{equation}
where $c_i$ and $c_a^{\dag}$ are the usual annihilation and creation operators which annihilates an electron in the occupied spinorbital $i$ and creates an electron in the vacant spinorbital $a$, respectively.
From here on, $i$, $j$, \ldots~are occupied spinorbitals, $a$, $b$, \ldots~denote virtual (unoccupied) spinorbitals, and $p$, $q$, $r$, and $s$ indicate arbitrary (orthonormal) spinorbitals.

The Schr\"odinger equation then reads
\begin{equation}
	\hH e^{\hT} \ket{\Phi} = E e^{\hT} \ket{\Phi},
\end{equation}
which can be rewritten as 
\begin{equation}
	\bH \ket{\Phi} = E \ket{\Phi}
\end{equation}
by defining the effective (non-Hermitian) similarity-transformed Hamiltonian 
\begin{equation}
	\label{eq:bH}
	\bH = e^{-\hT} \hH e^{\hT}.
\end{equation}
Although $\bH$ is non-Hermitian, the similarity transformation \eqref{eq:bH} ensures that $\bH$ has an energy spectrum that is identical to the original Hermitian operator $\hH$.
Besides, the exponential structure of the wave operator ensures rigorous size-extensivity and is responsible for the comparative high accuracy of the theory at relatively low computational cost.
The cluster amplitudes $t_{ij\dots}^{ab\dots}$ defined in Eq.~\eqref{eq:T_k} are the quantities to determine.

Truncating $\hT$ to double excitations, \ie, $\hT = \hT_1 + \hT_2$ with 
\rev{
\begin{subequations}
\begin{align}
	\hT_1 & = \sum_{ia} t_{i}^{a} c_a^{\dag} c_i,
        \\
	\hT_2 & = \frac{1}{4} \sum_{ijab} t_{ij}^{ab} c_a^{\dag} c_b^{\dag} c_j c_i,
\end{align}
\end{subequations}
}
%\begin{subequations}
%\begin{align}
%	\hT_1 & = \sum_{ia} t_{i}^{a} c_a^{\dag} c_i,
%	\\
%       \hT_2 & = \sum_{ijab} t_{ij}^{ab} c_a^{\dag} c_b^{\dag} c_j c_i,
%\end{align}
%\end{subequations}
defines CC with singles and doubles (CCSD) and one gets the single and double amplitudes, $t_{i}^{a}$ and $t_{ij}^{ab}$, via the amplitude equations
\begin{subequations}
\begin{align}
	\label{eq:T1_eq}
	\mel{\Phi_{i}^{a}}{\bH}{\Phi} & = 0,
	\\
	\label{eq:T2_eq}
	\mel{\Phi_{ij}^{ab}}{\bH}{\Phi} & = 0.
\end{align}
\end{subequations}
The (non-variational) CCSD energy is evaluated by projection,
\begin{equation}
	\ECC 
	= \mel{\Phi}{\bH}{\Phi} 
	= \frac{ \mel{\Phi}{ e^{-\hT} \hH e^{\hT} } {\Phi} }{ \mel{\Phi}{ e^{-\hT} e^{\hT} }{\Phi} },
\end{equation}
in contrast to its variational analog
\begin{equation}
	\label{eq:VCC}
	\EVCC 
	= \frac{\mel{\Phi}{e^{\hT^{\dag}} \hH e^{\hT}}{\Phi}}{\mel{\Phi}{e^{\hT^{\dag}} e^{\hT}}{\Phi}} 
	\ge \EFCI,
\end{equation}
(where the Rayleigh-Ritz variational principle has been used to determine the energy and the amplitudes) which is a true upper bound to the (exact) full configuration interaction (FCI) energy $\EFCI$. \cite{Bartlett_1988,VanVoorhis_2000}
Unfortunately, VCC is computationally intractable.
Indeed, even for truncated CC methods such as CCSD, VCC has factorial complexity since the power series expansion of the VCC energy \eqref{eq:VCC} does not naturally truncate, \ie, does not terminate before the $N$-electron limit.

%%%%%%%%%%%%%%%%%%%%%%%%%%%%%%%%%%%%%%%%%%%%%%%%%%
\section{Coupled cluster for excited states}
\label{sec:ESCC}
%%%%%%%%%%%%%%%%%%%%%%%%%%%%%%%%%%%%%%%%%%%%%%%%%%
Excited states can be attained via the equation-of-motion (EOM) formalism \cite{Rowe_1968,Monkhorst_1977,Koch_1990,Stanton_1993,Koch_1994} which consists in diagonalising the $\bH$ matrix in the space of excited determinants. 
If restricted to singly- and doubly-excited configurations, one obtains the EOM-CCSD method.
Loosely speaking, EOM-CCSD can be seen as a configuration interaction (CI) with singles and doubles (CISD) using $\bH$ instead of $\hH$. 
However, because $\bH$ is not Hermitian, its matrix representation is therefore non-symmetric, unlike the corresponding CI Hamiltonian matrix.
EOM-CCSD accurately describes single excitations \cite{Loos_2018b,Loos_2020c} but struggles to model excited states with strong double excitation character due to the lack of triples and higher excitations. \cite{Loos_2019c,Loos_2020d}
This issue can be cured by adding higher excitations but at a significant computational cost. \cite{Kucharski_1991,Christiansen_1995b,Kucharski_2001,Kowalski_2001,Hirata_2000,Hirata_2004}

In this paper, inspired by several groups, \cite{Piecuch_2000,Mayhall_2010,Lee_2019} we will focus on an alternative to EOM-CC and target excited states within the ``ground-state'' CC formalism described in Sec.~\ref{sec:GSCC} by searching for higher-energy solutions of the conventional CC amplitude equations [see Eqs.~\eqref{eq:T1_eq} and \eqref{eq:T2_eq}].
Indeed, as illustrated below, the amplitude equations form a set of polynomial equations in the cluster amplitudes, and they are, by definition, highly non-linear.
Therefore, the standard ground-state CC solution resulting from the usual self-consistent iterative procedure is not the only solution to this set of equations.
Unfortunately, these higher roots of the CC equations (that we label non-standard in the following) are hardly attainable in practice and one must be very cautious when targeting these solutions.

There exist three main factors that, we believe, significantly influence the solution that is reached.
First, the set of orbitals used to build the reference Slater determinant $\ket{\Phi}$ is rather important.
Ground-state HF orbitals are usually employed but alternative choices are possible, and excited-state HF orbitals [obtained via the maximum overlap method (MOM) \cite{Gilbert_2008,Barca_2014,Barca_2018a,Barca_2018b} or more fancy algorithms \cite{Hait_2020,Levi_2020a,Levi_2020b} to avoid variational collapse] are getting more and more common. \cite{Lee_2019,Hait_2020,Carter-Fenk_2020,Hait_2021}
Second, the starting amplitudes that are usually derived from perturbation theory (CCSD guess amplitudes are usually taken as MP2 amplitudes) may influence the outcome of the iterative process.
Third, the type of iterative algorithms (usually based on the Newton-Raphson method and supplemented by Pulay's DIIS method \cite{Pulay_1980,Pulay_1982,Scuseria_1986}) must also be carefully chosen so as to target, for example, saddle points instead of minima.
Throughout the text, we refer to Newton-Raphson when the full Jacobian (for zeros) or Hessian (for extrema) matrix is employed, and to quasi-Newton when they are provided only approximately.

The seminal works of Zivkovic and Monkhorst were the first to shed some light on the existence conditions of the higher roots of the CC equations. \cite{Zivkovic_1977,Zivkovic_1978}
Adamowicz and Bartlett studied the attainability of some excited states of the \ce{LiH} molecule using the dominant determinants of CI expansions as a reference state for a CCSD calculation. \cite{Adamowicz_1985} 
Later, Jankowski \textit{et al.}~investigated the CCD solutions of $^{1}A_1$ symmetry in the \ce{H4} molecule, \cite{Jankowski_1994,Jankowski_1994a,Jankowski_1995} evidencing that some non-standard CC solutions are unphysical.
Moreover, they showed that the number of attainable solutions depends on the choice of the reference determinants. \cite{Jankowski_1995}

A crucial step in the study of non-standard CC solutions was the introduction of the homotopy method by Kowalski \textit{et al.}, \cite{Kowalski_1998,Kowalski_1998a} which allows to find all the solutions of a set of polynomial equations. \cite{Verschelde_1994}
The gist of this method is to create an analytic continuation between a set of equations for which the solutions are known at $\lambda = 0$ and the set of equations to be solved at $\lambda=1$.
The key difficulty of the homotopy method is to be able to follow distinctly the solutions from $\lambda = 0$ to $\lambda=1$.
In practice, this ``path-tracking'' is very difficult and computationally expensive.
In a subsequent series of papers, Jankowski and Kowalski explored in more details the structure of the CC solutions using the same homotopy method \cite{Jankowski_1999,Jankowski_1999a,Jankowski_1999b,Jankowski_1999c} (see also Refs.~\onlinecite{Paldus_1993,Kowalski_2000,Kowalski_2000a}).
In the meantime, Piecuch and Kowalski published an extensive review along the same lines, \cite{Piecuch_2000} and we refer the interested reader to this instructive review for additional information.

Few years later, the homotopy method was used to study the Pariser-Parr-Pople model of benzene and [10]-annulene. \cite{Podeszwa_2002,Podeszwa_2003}
More recently, Mayhall \textit{et al.}~pointed out that the problem of the CC solution structure still needs to be addressed for real systems, and they investigated the appearance of multiple CCSD solutions for the \ce{NiH} molecule. \cite{Mayhall_2010}
Finally, Lee \textit{et al.}~targeted doubly-excited states and double core hole states of small molecules using orbital-optimized non-Aufbau determinants. \cite{Lee_2019}

%%%%%%%%%%%%%%%%%%%%%%%%%%
\section{Pair CCD for ground states}
%%%%%%%%%%%%%%%%%%%%%%%%%%
Our primary goal here is to investigate precisely the type and nature of excited states that one can reach (and their actual number) in the simple case of pair CCD (pCCD), which provides a reasonable description of strong correlation for a wide variety of systems. 
%\cite{Henderson_2014a,Henderson_2014b,Stein_2014,Gomez_2016,Shepherd_2016,Boguslawski_2017a,Boguslawski_2017b,Boguslawski_2019}
\cite{Henderson_2014a,Henderson_2014b,Stein_2014,Gomez_2016,Shepherd_2016,Boguslawski_2016a,Boguslawski_2016b,Boguslawski_2017a,Boguslawski_2017b,Boguslawski_2019}
pCCD, which was first named the antisymmetric product of 1-reference orbital geminals, 
%\cite{Limacher_2013,Limacher_2014,Tecmer_2014,Boguslawski_2014a,Boguslawski_2014b,Boguslawski_2014c,Tecmer_2015,Boguslawski_2015,Boguslawski_2016a,Boguslawski_2016b} 
\cite{Limacher_2013,Limacher_2014,Tecmer_2014,Boguslawski_2014a,Boguslawski_2014b,Boguslawski_2014c,Tecmer_2015,Boguslawski_2015}
restricts CCD to the seniority zero subspace, the subspace of all closed-shell determinants. \cite{Bytautas_2011}  
In the present context of closed-shell systems, the seniority number is defined as the number of unpaired electrons in a determinant. \cite{Ring_1980}
While expanding the wave function in terms of the excitation rank has been proved to be slowly convergent for strongly correlated systems, the seniority zero subspace seems to be efficient at describing such systems. \cite{Bytautas_2011}  
Because the pCCD energy is not invariant with respect to orbital rotations, the orbitals must be optimized to enhance the amount of correlation energy recovered. \cite{Bytautas_2011,Stein_2014,Limacher_2014}

pCCD has quite interesting features because it provides, at mean-field computational cost (disregarding the cost of the two-electron integral transformation), very similar ground-state energies as doubly-occupied CI (DOCI), a method with formal exponential scaling. \cite{Allen_1962,Smith_1965,Veillard_1967,Weinhold_1967,Couty_1997,Kollmar_2003,Bytautas_2011}
This surprising observation has been shown to hold for both canonical HF orbitals and energetically-optimized orbitals.
(Like pCCD, DOCI is not invariant to the orbitals with respect to which seniority is defined.)
However, the equivalence between pCCD and DOCI is not strict mathematically speaking \cite{Henderson_2014a,Henderson_2015,Shepherd_2016} and significant differences have been revealed for pairing Hamiltonians in particular. \cite{Henderson_2014b}
Here, we propose to investigate whether or not this energetic similarity between pCCD and DOCI pertains for excited states in the case of molecular systems.

The pCCD equations can be easily obtained from the usual CCD equations by restricting the excitation manifold to electron pairs, \ie, 
\begin{equation}
	\hat{T} = \sum_{ia} t_{ii}^{aa} P_a^{\dag} P_i,
\end{equation}
where we have defined, for convenience, the pair operators $P_q^{\dag} = c_{q\uparrow}^{\dag} c_{q\downarrow}^{\dag}$.
For the sake of conciseness, we denote the pair amplitudes $t_{ii}^{aa}$ as $t_i^a$ from here on, as we will not consider single excitations.
By considering the similarly-transformed Hamiltonian $\bH$ defined in Eq.~\eqref{eq:bH}, we obtain 
\begin{subequations}
\begin{align}
	E & = \mel{\Phi}{\bH}{\Phi},
	\\ 
	r_{i}^a & = \mel{\Phi}{P_i^\dag P_a \bH}{\Phi} = 0,
\end{align}
\end{subequations}
which yields the following explicit equations \cite{Henderson_2014a}
\begin{subequations}
\begin{align}
	\label{eq:EpCCD}
	E & = \mel{\Phi}{\hat{H}}{\Phi} + \sum_{ia} t_i^a \ERI{ii}{aa},
	\\ 
	\label{eq:T_eq_pCCD}
	\begin{split}
	r_{i}^a 
	& = \ERI{ii}{aa} 
	+ 2 (f_{a}^{a} - f_{i}^{i} - \sum_j \ERI{jj}{aa} t_j^a - \sum_b \ERI{ii}{bb} t_i^b ) t_i^a
	\\
	& - 2 (2 \ERI{ia}{ia} - \ERI{ia}{ai} + \ERI{ii}{aa} t_i^a) t_i^a 
	\\
	& + \sum_b \ERI{aa}{bb} t_i^b + \sum_j \ERI{ii}{jj} t_j^a + \sum_{jb} \ERI{jj}{bb} t_j^a t_i^b,
	\end{split}
\end{align}
\end{subequations}
where $f_{p}^{q}$ are elements of the Fock operator and $\ERI{pq}{rs} = \braket{pq}{rs}$ are two-electron integrals in the spatial orbital basis written following Dirac's notation.

Equation \eqref{eq:T_eq_pCCD} provide $OV$ quadratic equations in the $OV$ unknown amplitudes $t_i^a$ (where $O$ and $V$ are the numbers of occupied and virtual spatial orbitals respectively).
An upper bound for the number of solutions of a system of polynomial equations is provided by B\'ezout's number which is equal to $2^{OV}$ in the present case. \cite{HartshorneBook,Kowalski_1998a,Jankowski_1999a,Piecuch_2000,Burton_2018}
In contrast the number of DOCI solutions is strictly equal to the binomial coefficient $\mqty(O+V\\O)$.

Starting with amplitudes borrowed from second-order ``pair'' M{\o}ller-Plesset
\begin{equation}
	\Tilde{t}_i^a = \frac{\ERI{ii}{aa}}{2f_{a}^{a} - 2f_{i}^{i}},
\end{equation}
the usual approach to solve these equations employs a quasi-Newton algorithm (where the differences of the Fock diagonal elements are taken as an approximate Jacobian matrix) which consists in updating the pCCD amplitudes as
\begin{equation}
	\label{eq:t_update_1}
	t_i^a \leftarrow t_i^a - \frac{r_i^a}{2f_{a}^{a} - 2f_{i}^{i}},
\end{equation}
where the pCCD residuals $r_i^a$ are given by Eq.~\eqref{eq:T_eq_pCCD} and are only equal to zero at convergence.
These equations can be solved in cubic computational cost if one defines an intermediate array to bypass the only quartic step [see last term in Eq.~\eqref{eq:T_eq_pCCD}].
%\cite{Henderson_2014a}
\cite{Henderson_2014a,Boguslawski_2014a,Boguslawski_2014b,Boguslawski_2014c}

Unlike CCD, the pCCD energy depends on orbital rotations within the occupied and the virtual spaces (besides the occupied-virtual rotations).
By variationally optimizing all orbitals, we have what may be called orbital optimized pCCD (oo-pCCD).

For the sake of consistency, we briefly review how to perform orbital optimization \cite{Scuseria_1987,Bozkaya_2011} for a pCCD ansatz. \cite{Henderson_2014a}
To do so, the energy has to be expressed as a functional to be minimized, \ie, 
\rev{
\begin{equation}
        \label{eq:energy_functional}
	\tilde{E} = \mel{\Phi}{(1+\hat{Z}) \bH} {\Phi},
\end{equation}
}
%\begin{equation}
%        \label{eq:energy_functional}
%        \tilde{E} = \mel{\Phi}{(1+Z) \bH} {\Phi},
%\end{equation}
where the de-excitation operator
\begin{equation}
        \label{eq:Z_operator}
        \hat{Z} = \sum_{ia} z_a^i P_i^{\dag} P_a
\end{equation}
is introduced.
Then, imposing the functional to be stationary with respect to the $z$-amplitudes, \ie, $\partial \tilde{E}/\partial z_a^i = 0$, immediately returns the equations for the $t$-amplitudes [Eq.~\eqref{eq:T_eq_pCCD}], and thereby for the pCCD energy [Eq.~\eqref{eq:EpCCD}].
Doing the same for the $t$-amplitudes, \ie, $\partial \tilde{E}/\partial t_i^a = 0$, gives a new set of (linear) equations to be solved for the $z$-amplitudes (see Ref.~\onlinecite{Henderson_2014a} for its explicit expression).

As commonly done, the orbital rotations are introduced by an exponential unitary operator, $e^{\hat{\kappa}}$, which acts on the right- and left-hand wave functions. \cite{Helgakerbook}
The operator 
\begin{equation}
        \label{eq:kappa}
	\hat{\kappa} = \sum_{p>q} \kappa_{pq} ( c_{p\uparrow}^{\dag}   c_{q\uparrow}^{}   - c_{q\uparrow}^{\dag}   c_{p\uparrow}^{} 
					      + c_{p\downarrow}^{\dag} c_{q\downarrow}^{} - c_{q\downarrow}^{\dag} c_{p\downarrow}^{} )
\end{equation}
encompasses all unique orbital rotations and its anti-hermiticity guarantees the unitarity of $e^{\hat{\kappa}}$, hence the orthogonality of the rotated orbitals.
Next, the energy is expressed as a functional of the orbital rotation operator $\hat{\kappa}$, \ie,
\begin{equation}
        \label{eq:energy_functional_kappa}
	\tilde{E}(\hat{\kappa}) = \mel{\Phi}{(1+Z) e^{-\hT} e^{-\hat{\kappa}} \hH e^{\hat{\kappa}} e^{\hT} } {\Phi}.
\end{equation}
Using matrix representations, stationary points with respect to $\kappa_{pq}$ can be found with the Newton-Raphson method which consist in expanding the energy to second order around $\bm{\kappa}=\bm{0}$
\begin{equation}
	\label{eq:energy_expansion}
	\tilde{E}(\bm{\kappa}) \approx \tilde{E}(\bm{0}) + \bm{g} \cdot \bm{\kappa} + \frac{1}{2} \bm{\kappa^{\dag}} \cdot \bm{H} \cdot \bm{\kappa},
\end{equation}
where $\bm{g}$ is the orbital gradient and $\bm{H}$ is the orbital Hessian, both evaluated at $\bm{\kappa}=\bm{0}$, \ie,
\begin{align}
	g_{pq} & = \left. \pdv{\tilde{E}(\bm{\kappa})}{ \kappa_{pq}} \right|_{\bm{\kappa} = \bm{0}},
	&
	H_{pq,rs} & = \left. \pdv{\tilde{E}(\bm{\kappa})}{ \kappa_{pq}}{ \kappa_{rs}} \right|_{\bm{\kappa} = \bm{0}}.
\end{align}
The approximated energy functional is minimized with the orbital rotation $\bm{\kappa} = - \bm{H}^{-1} \cdot \bm{g} $, which then defines a new second-order approximation.
This procedure is repeated until the orbitals become stationary, \ie, $\norm{\bm{g}}_\infty < \tau$, where $\tau$ is a user-defined threshold which has been set to $10^{-5}$ a.u.~in the present study.

All the additional equations required to energetically optimize the orbitals can be found in Ref.~\onlinecite{Henderson_2014a}, in particular the one- and two-body density matrices required to compute the orbital gradient and Hessian matrices.

%%%%%%%%%%%%%%%%%%%%%%%%%%
\section{Pair CCD for excited states}
%%%%%%%%%%%%%%%%%%%%%%%%%%
Excited states can be accessed with a pCCD reference via the EOM formalism as investigated by Boguslawski. \cite{Boguslawski_2016b,Boguslawski_2017b,Boguslawski_2019}
By including single and paired double excitations in the EOM excitation operator,
one arrives at the EOM-pCCD+S model. \cite{Boguslawski_2016b,Boguslawski_2017b}
Approximately accounting for the nonpair excitations provides a more sophisticated approach, the EOM-pCCD-LCCSD method. \cite{Boguslawski_2019}
More precisely, the reference is described with a hybrid pCCD and linearized CC approach for the nonpair excitations,
while all single and double excitations are included in the EOM excitation operator.
Both methods have been used for computing excitation energies of singly- and doubly-excited states, 
%\cite{Boguslawski_2016b,Boguslawski_2017b,Boguslawski_2019,Nowak_2019}
\cite{Boguslawski_2016b,Boguslawski_2017b,Boguslawski_2019,Nowak_2019,Tecmer_2019}
and very good performance was attained in these applications.
However, excitation energies tend to be somewhat overestimated with respect to reference (FCI) values.
This takes place because the reference wave functions is built from orbitals optimized for the ground state (for either HF or pCCD wave function),
thus biasing the calculations towards this state.

One of our main goals here is to explore an alternative route for describing doubly-excited states, while still making use of the pCCD ansatz.
The idea is to perform independent, state-specific oo-pCCD calculations for the ground state and for a specifically targeted doubly-excited state.
\rev{This defines the $\Delta$oo-pCCD method, where excitation energies are evaluated from the energy difference between these two separate oo-pCCD calculations.}
In this way, we hope that our $\Delta$oo-pCCD method can provide a more balanced description of correlation effects for both states,
and thus more accurate excitation energies when compared with more computationally demanding CC alternatives, such as CC3 and EOM-CCSDT.

%%%%%%%%%%%%%%%%%%%%%%%%%%%%%%%%
\section{Computational details}
\label{sec:compdet}
%%%%%%%%%%%%%%%%%%%%%%%%%%%%%%%%

A selected version of DOCI was implemented in {\QP} via a straightforward modification of the \textit{configuration interaction using a perturbative selection made iteratively} (CIPSI) algorithm \cite{Huron_1973,Giner_2013,Giner_2015} where only the seniority zero determinants of the FCI space are considered. \cite{Garniron_2017b,Garniron_2018,Garniron_2019}
(Note that the calculation of the second-order perturbative correction is also restricted to the seniority zero subspace. \cite{Garniron_2017b})
A similar modification has been employed by Shepherd \textit{et al.} \cite{Shepherd_2016} to perform DOCI calculations within FCIQMC. \cite{Booth_2009}
\rev{In our DOCI implementation, the roots are located with the standard Davidson diagonalization method. \cite{Davidson_1975}}
The pCCD method and the corresponding orbital optimization algorithm was also implemented in {\QP} following Ref.~\onlinecite{Henderson_2014a}.
The FCI calculations presented here are also performed with the CIPSI algorithm implemented in {\QP}. \cite{Huron_1973,Giner_2013,Giner_2015,Garniron_2019}
Solving the CC equations and optimizing the orbitals for excited states are central to our discussion and are thus discussed in separate sections below.
Details regarding the particular applications are also given in their respective sections.
Complementary CC3, EOM-CCSDT, and EOM-CCSDTQ calculations were also performed with the CFOUR package. \cite{Matthews_2020}

%%%%%%%%%%%%%%%%%%%%%%%%%%%%%%%%
\section{Results and discussion}
\label{sec:res}
%%%%%%%%%%%%%%%%%%%%%%%%%%%%%%%%

%================================
\subsection{Targeting excited states}
\label{subsec:He_exc}
%================================

When aiming at excited states, important aspects regarding the algorithms for solving the pCCD equations and the orbital optimization should be addressed first.
We illustrate these points with pCCD calculations for the helium atom in the small 6-31G basis set made of two basis functions,
where the working equations can be solved analytically for the single amplitude $t \equiv t_{h}^l$ between the HOMO ($h$) and the LUMO ($l$) orbitals.

\begin{figure}[h!]
\includegraphics[width=\linewidth]{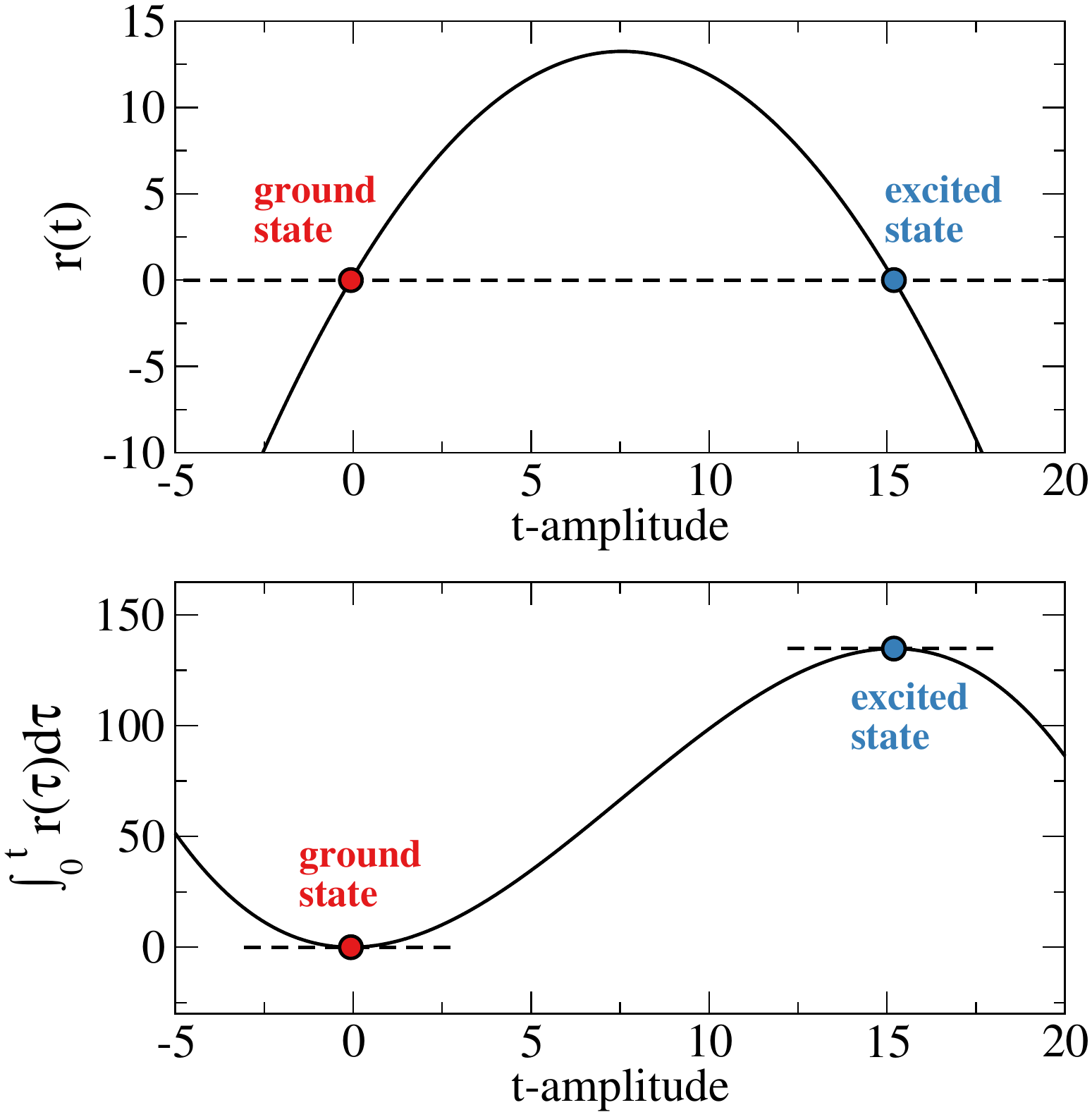}
        \caption{Ground and doubly-excited states for the helium atom (6-31G basis set), identified as the roots of the pCCD residual polynomial (top panel),
        or as the stationary points of the integrated residual polynomial (bottom panel).}
        \label{fig:scan_t}
\end{figure}

The first aspect concerns the updating step of the algorithm [see Eq.~\eqref{eq:t_update_1}], which must be modified to properly target excited states.
Once the two canonical HF orbitals are obtained, the pCCD amplitude $t$ is obtained by finding the roots of a single second-order polynomial equation
\begin{equation}
        \label{eq:residual_1}
        r(t) = \ERI{hh}{ll} + \left( 2 f_l^l - 2 f_h^h - 4 \ERI{hl}{hl} + 2 \ERI{lh}{hl} + \ERI{ll}{ll} + \ERI{hh}{hh} \right) t - \ERI{ll}{hh} t^2,
\end{equation}
which is shown in Fig.~\ref{fig:scan_t} (top panel).
The first root is located at small $t$ and corresponds to the ground state (where the HOMO orbital is doubly occupied), whereas the second root appears at a larger $t$ value and corresponds to the doubly-excited state (where the LUMO is doubly occupied).
As readily seen, Eq.~\eqref{eq:t_update_1} implicitly assumes that the derivative of the residual with respect to $t$ (the denominator) can be approximated by the orbital energy differences ($2 f_l^l - 2 f_h^h$ in our case study).
However, this quantity is independent of the amplitudes (and usually positive for a ground-state reference). 
This assumption clearly breaks down far from the ground-state solution.
At a given iteration, $t$ will keep decreasing/increasing when the residual is positive/negative. 
Therefore, one either converges to the ground state (when the guess amplitude is smaller than the second root), or diverges (when the guess amplitude is larger than the second root).

%%% FIG 1 %%%
%%% %%% %%%

The picture might become more familiar when the root-finding problem is framed as an optimization problem, also depicted in Fig.~\ref{fig:scan_t} (bottom panel) for our case study.
In this framework, we are looking for the stationary points, but the equivalent of Eq.~\eqref{eq:t_update_1} only works properly when looking for minima.
The equivalent assumption is that the second derivative is constant and positive, even though it is actually negative close to the maximum.
Locating this additional stationary point thus requires information about the actual curvature.

Back to the original root finding problem, this means employing the first derivative of the residual as the denominator in Eq.~\eqref{eq:t_update_1}.
Doing so is precisely the Newton-Raphson method.
For the single amplitude case, the correct residual derivative is
\begin{equation}
        \label{eq:derivative_residual_1}
        \dv{r(t)}{t} = 2 f_l^l - 2 f_h^h - 4 \ERI{hl}{hl} + 2 \ERI{lh}{hl} + \ERI{ll}{ll} + \ERI{hh}{hh} - 2 \ERI{ll}{hh} t,
\end{equation}
while in Eq.~\eqref{eq:t_update_1} only the orbital energy differences ($2 f_l^l - 2 f_h^h$) are considered.
When the terms involving the two-electron integrals are accounted for,
both roots depicted in Fig.~\ref{fig:scan_t} can be located, each one with a well-defined basin of attraction.

Therefore, the usual updating algorithm can only find the roots where all the residual first derivatives are positive.
While this holds in general for the lowest-lying solution, it does not for higher-lying roots.
In the latter case, one should provide the Jacobian matrix, or at least some descent approximation of it.
Here we have employed the Newton-Raphson method, by evaluating and inverting the full Jacobian matrix $\bm{J}$ of the system of residual equations [see Eq.~\eqref{eq:T_eq_pCCD}], \ie, $J_{ia,jb} = \partial r_i^a / \partial t_j^b$, which in pCCD is given as
\begin{align}
        \label{eq:jacobian}
        \begin{split}
        J_{ia,jb} & = \left[ 2 (f_{a}^{a} - f_{i}^{i}) - 4 \ERI{ia}{ia} + 2 \ERI{ia}{ai} \right] \delta_{ij} \delta_{ab} \\
                  & + \left[ \ERI{aa}{bb} - \ERI{jj}{aa} t_i^a + (1-2\delta_{ab}) \sum_{k\neq i} \ERI{kk}{bb} t_k^a \right] \delta_{ij} \\
                  & + \left[ \ERI{ii}{jj} - \ERI{ii}{bb} t_i^a + (1-2\delta_{ij}) \sum_{c\neq a} \ERI{jj}{cc} t_i^c \right] \delta_{ab}.
        \end{split}
\end{align}
Then, at each Newton-Raphson step, the pCCD amplitudes are updated as
\begin{equation}
        \label{eq:t_update_2}
	t_i^a \leftarrow t_i^a - \sum_{jb} (\bm{J}^{-1})_{ia,jb} r_j^b.
\end{equation}
Compared with the usual quasi-Newton approach [see Eq.~\eqref{eq:t_update_1}], the main overhead concerns storing and inverting the Jacobian matrix.
On the other hand, its computation comes with a minor cost, as all the needed contractions are already performed for the residuals.
Alternatively, we may approximate the Jacobian matrix by its diagonal, which requires much less memory.
Making use of the full or diagonal Jacobian matrix proved to be quite reliable when targeting excited states, besides requiring much fewer iterations to converge.

\subsection{Orbital optimization}
\label{subsec:He_opt}

\begin{figure}[h!]
\includegraphics[width=\linewidth]{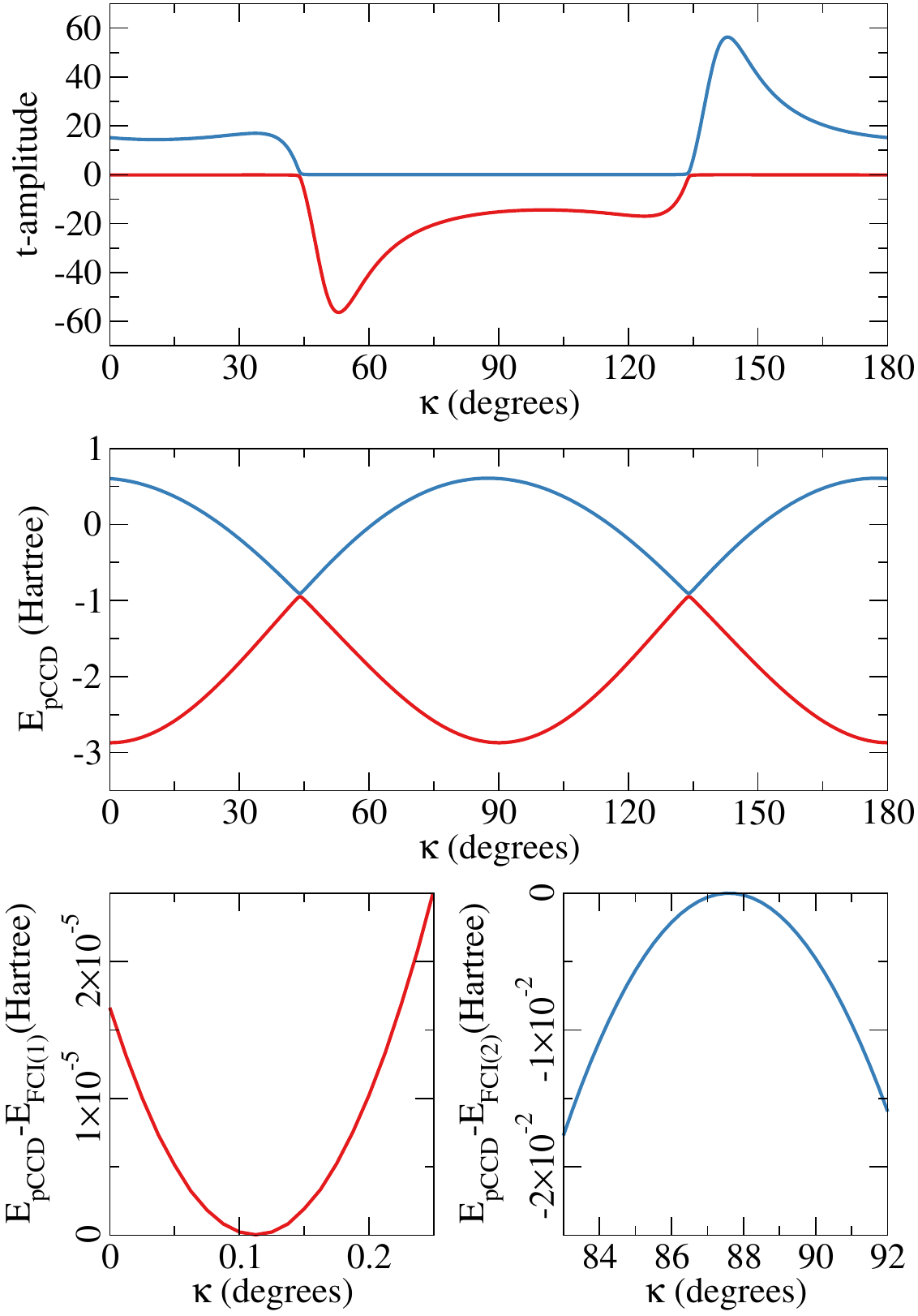}
        \caption{Amplitudes (top), pCCD total energies $E_\text{pCCD}$ (middle), and its difference to the first [$E_\text{FCI(1)}$] and second [$E_\text{FCI(2)}$] FCI energies (bottom) for the lower (red) and higher (blue) pCCD solutions of the helium atom computed with the 6-31G basis set along the single orbital rotation parameter $\kappa$ between HOMO and LUMO HF orbitals ($\kappa = 0$ represents the ground-state HF reference).
	}
        \label{fig:scan_kappa}
\end{figure}

The orbital optimization procedure represents the second key aspect to address when targeting excited states.
Still considering the helium atom in the 6-31G basis set, we present in Fig.~\ref{fig:scan_kappa} how the pCCD amplitudes and energies behave as one varies the reference orbitals, which are solely determined by
the parameter $\kappa$ that rotates the HOMO and LUMO HF orbitals.
(The case depicted in Fig.~\ref{fig:scan_t} corresponds to $\kappa=0$, where the HOMO HF orbital is doubly occupied.)
Crossings between the two pCCD solutions are observed around 44 and 134 degrees for both $t$-amplitudes and energies.
In between these points, the reference wave function resembles more the excited state, while the ground state is reached with large $t$-amplitudes.
The bottom panels of Fig.~\ref{fig:scan_kappa} highlight the stationary points for the pCCD energy, where the reference orbitals are said to be optimized.
In particular, oo-pCCD and FCI deliver the same energies, as it should for a two-electron system. \cite{Henderson_2014a}
For the ground state, this takes place with very slight orbital mixing ($\kappa = 0.11$ degrees),
while optimized orbitals for the doubly-excited state are found at $\kappa = 87.6$ degrees, close to double occupation of the LUMO HF orbital ($\kappa = 90$ degrees).
In contrast, a more significant mixing of the HF orbitals provides a very poor reference (for either state), and pCCD cannot recover from that.

%%% FIG 2 %%%
%%% %%% %%% 

For helium, obtaining oo-pCCD solutions is relatively straightforward.
However, as the number of orbital rotation parameters increases, one needs a reliable and hopefully black-box optimization protocol.
Optimizing orbitals for excited states has proved to be considerably more challenging than for the ground state.
While at least one minimum exists for the latter, the former typically appear as saddle points.
Additionally, a multitude of local minima and saddle points can be expected, and a correspondence between each stationary point and each physical state does not necessarily exist.
One might encounter more than one stationary point that actually represents the same physical state.
Alternatively, some of them could be artifacts of the underlying (approximate) level of theory.
Furthermore, if the optimization is aimed at a given excited state, there is no way to tell beforehand the order of the corresponding saddle point, not even if it exists.
And even when we do land in such a point, there is no guarantee that it is the only one describing the targeted state.
Finally, one might converge to the ground state when another state was intended,
and this collapse to the wrong solution should be avoided when implementing a robust algorithm.

Our previous discussion about how to optimize $t$-amplitudes for excited states [see Sec.~\ref{subsec:He_exc}] applies in the same way for orbital optimization.
When higher-lying pCCD solutions are targeted, some information about the orbital rotation Hessian has to be provided (even if approximate).
Here, we have computed the full orbital Hessian and gradients~\cite{Henderson_2014a} during the entire optimization process.
We also tried computing only the diagonal Hessian, but convergence deteriorated significantly.
Evaluating and storing the full Hessian is affordable for the cases we have considered here, and we have proceeded as such.
We further explored the DIIS algorithm, \cite{Pulay_1980}
but that was often unstable for excited states, or favored the collapse to the ground state.
\rev{We stress, however, that the orbital optimization protocol that we have devised here (for either ground or excited states) is computationally feasible only for small molecules.
In view of the need to compute and store the full Hessian, our approach would become impractical for larger systems, where more approximate and efficient algorithms should be employed instead.
Tailored algorithms have recently allowed orbital optimization within pCCD for ground-state calculations of a model vitamin B12 compound, \cite{Boguslawski_2021}
and adapted versions of such algorithms might be required for oo-pCCD calculations targeting excited states in large systems.}

The orbitals were optimized with a modified Newton-Raphson method.
At each iteration, the full Hessian is diagonalized and before solving the corresponding linear system, the eigenvalues are modified as follows.
The positive ones are increased by a constant positive factor, which effectively damps the next step along the corresponding eigenvector direction.
Doing the same for the negative eigenvalues could turn one positive,
thereby changing the Hessian structure and guiding the optimization toward a stationary point with an unintended saddle order.
Therefore, we have added a constant negative factor to the negative eigenvalues, which damps the step while preserving the Hessian structure.
When there are more negative eigenvalues than intended, we step along the gradients corresponding to the largest negative ones.
By carefully choosing the damping factors, we were able to converge to a desired stationary point for the states we have targeted here.

At this point we summarize the complete algorithm employed in our oo-pCCD calculations.
Each calculation starts with ground-state HF orbitals, and when doubly-excited states are concerned, the corresponding non-Aufbau occupancy is employed.
The pCCD equations are solved for the $t$-amplitudes [Eq.~\eqref{eq:T_eq_pCCD}] with the Newton-Raphson algorithm.
With these converged amplitudes, the de-excitation $z$-amplitudes are obtained with a single Newton-Raphson step (since they appear linearly~\cite{Henderson_2014a}).
Both sets of amplitudes, as well as the one- and two-electron integrals are needed to compute one- and two-body density matrices.
These, in turn, are used to compute orbital gradient and Hessian, which provide the orbital rotation parameters according to our modified Newton-Raphson algorithm.
This defines a new reference wave function, and the process repeats iteratively until convergence.

%================================
\subsection{Hydrogen chains}
%================================

As a first example, we consider the linear \ce{H4} molecule in a minimal basis (STO-6G), and we compute the ground- and excited-state energies of this system at the pCCD and DOCI levels,
as a function of the distance between the (equally-spaced) hydrogen atoms $R_{\ce{H-H}}$.
This corresponds to a system with 4 electrons in 4 spatial orbitals with respective symmetries $\sigma_g$, $\sigma_u$, $\sigma_g^*$, and $\sigma_u^*$ (in ascending energies).

\begin{figure}
        \includegraphics[width=\linewidth]{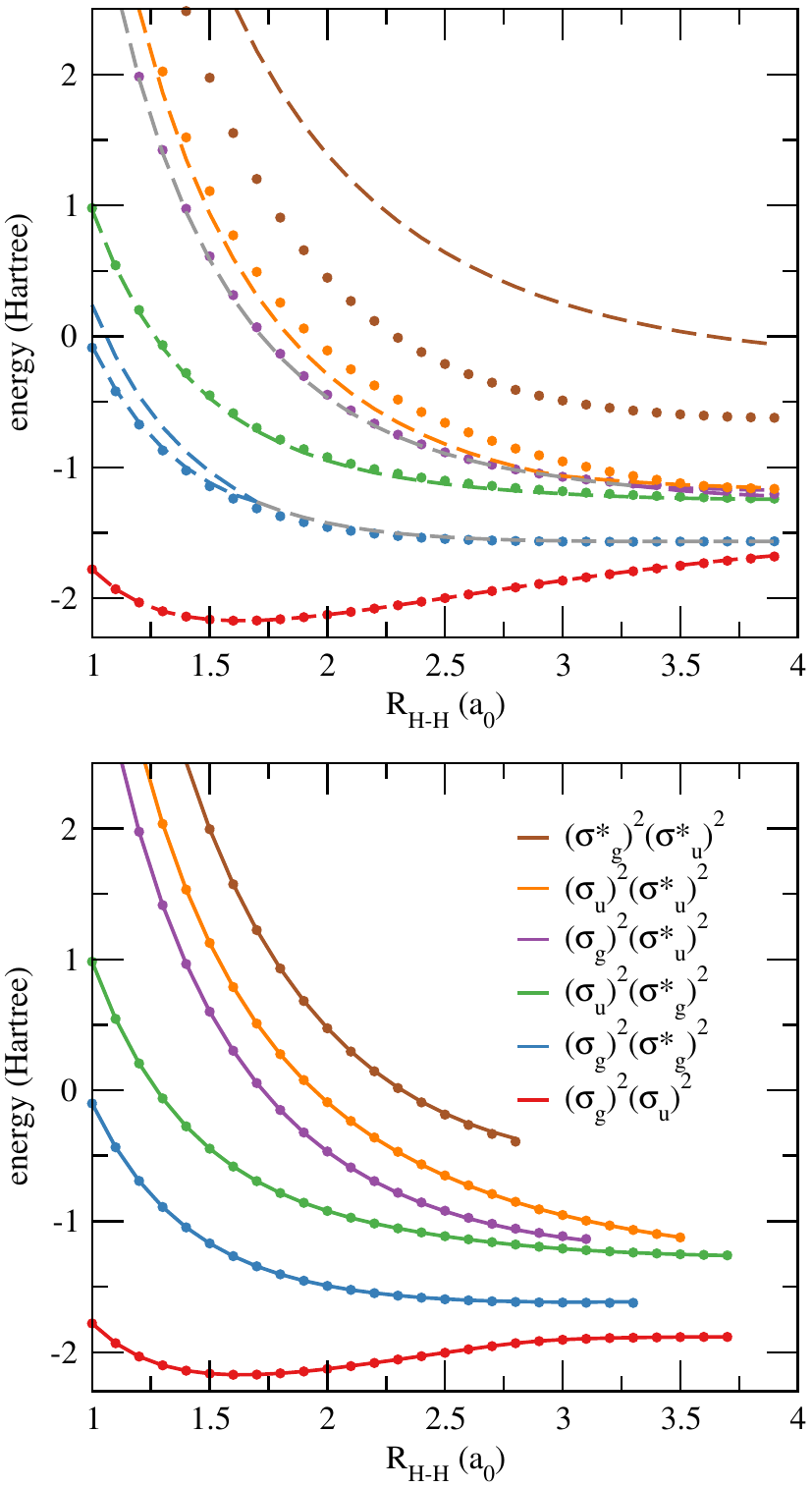}
        \caption{Ground- and excited-state energies of the linear \ce{H4} molecule in the STO-6G basis obtained with various methods, 
	as functions of the distance between the (equally-spaced) hydrogen atoms $R_{\ce{H-H}}$.
	On the top panel, pCCD (dashed lines) and DOCI (markers) energies, both computed with ground-state HF orbitals.
	(Grey dashed lines denote the real part of the complex solutions.)
	On the bottom panel, oo-pCCD (solid lines) and DOCI (markers) energies, both computed with state-specific optimized orbitals.
	Raw data are provided in the {\SupInf}.
        \label{fig:H4_both}}
\end{figure}

\begin{figure}
        \includegraphics[width=\linewidth]{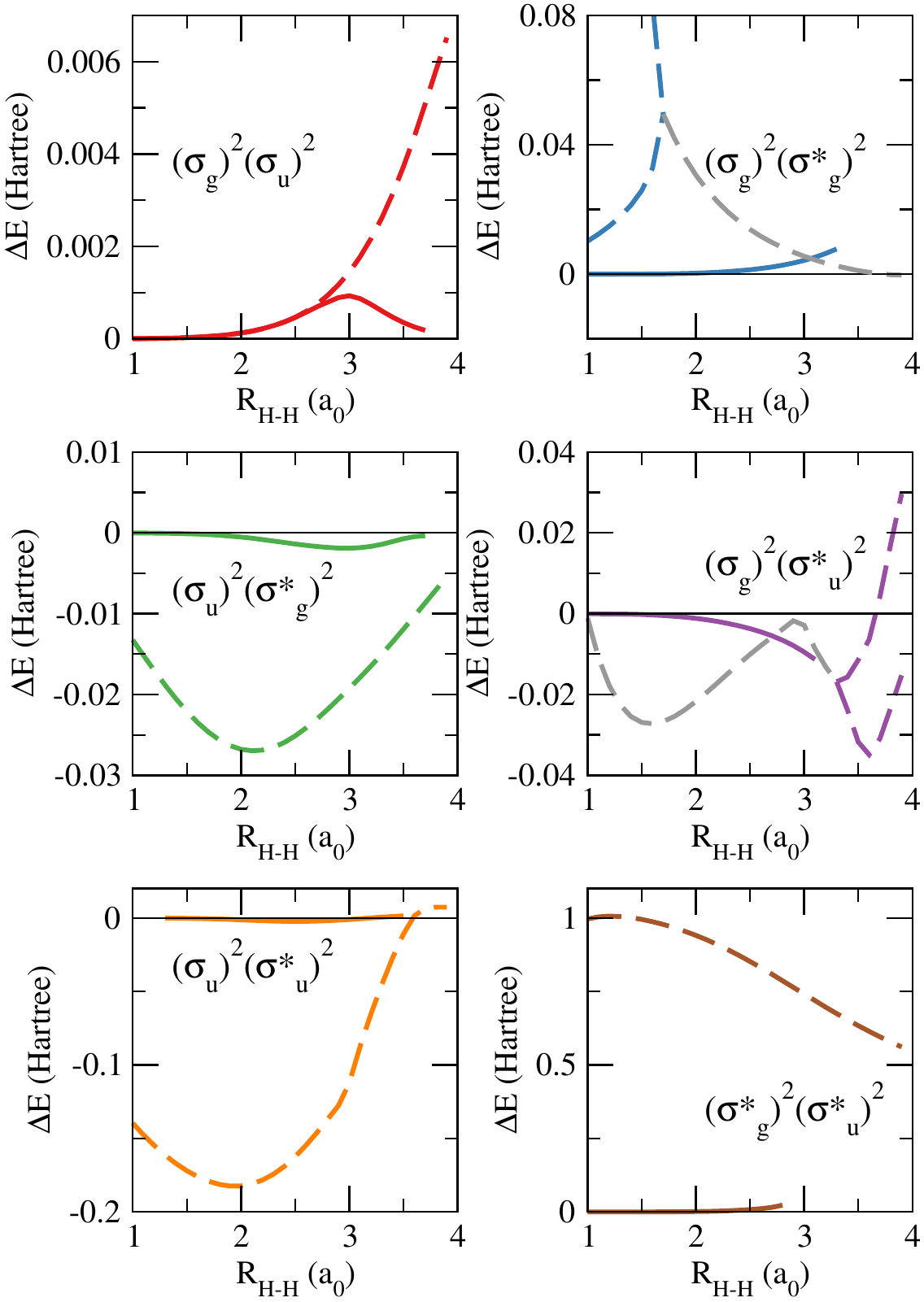}
	\caption{Differences between oo-pCCD and DOCI energies, both computed with ground-state HF (dashed lines) and state-specific optimized orbitals (solid lines),
	for the linear \ce{H4} molecule in the STO-6G basis,
	as functions of the distance between the (equally-spaced) hydrogen atoms $R_{\ce{H-H}}$.
	Grey dashed lines correspond to the regions where the pCCD solutions become complex (the real part is used to compute the difference).
        \label{fig:H4_diffs}}
\end{figure}

We have considered two scenarios. In the first (top panel of Fig.~\ref{fig:H4_both}), canonical HF orbitals were employed throughout, 
and both pCCD and DOCI potential energy curves represent different solutions of their corresponding equations. 
In the second scenario (bottom panel of Fig.~\ref{fig:H4_both}), oo-pCCD calculations were performed for each targeted state, 
and the same set of optimized orbitals were used to find the matching DOCI root.
In order to help our discussion, the differences between pCCD and DOCI energies (with and without orbital optimization) are also shown in Fig.~\ref{fig:H4_diffs}.

%%% FIG 3 %%%
%%% %%% %%%

%%% FIG 4 %%%
%%% %%% %%%

\rev{The obtained pCCD optimized orbitals are shown in the {\SupInf}.
Even though they were allowed to break spatial symmetry, only symmetry-preserving orbitals were found for the excited states.
This does not constitute a limitation for our purpose of comparing the two methods, 
yet we might still be missing possible symmetry-broken oo-pCCD solutions. \cite{Fukutome_1981,Stuber_2003}}
In particular, the ground-state orbital Hessian has one negative eigenvalue between \SI{1.4}{\bohr} and \SI{2.7}{\bohr},
which indicates the existence of a lower-lying solution with symmetry-broken orbitals. \cite{Seeger_1977,Jimenez-Hoyos_2011,Henderson_2018,Burton_2021}
Similarly, the number of negative Hessian eigenvalues changes from 3 to 2 when stretching beyond 2.3--\SI{2.4}{\bohr}, for the second excited state,
and from 6 to 5 above 1.9--\SI{2.0}{\bohr}, for the fifth excited state.
In contrast, the index is constant for first, third, and fourth excited states (2, 3, and 3, respectively).
\rev{While symmetry-broken solutions describing these doubly-excited states might still exist, they were not found,
suggesting that the potential number of multiple solutions for this small basis set might not be too large.}

For the same set of orbitals, pCCD and DOCI ground-state energies are very close, as expected.
We notice, however, that orbital optimization improves the comparison at more stretched geometries (see top left panel of Fig.~\ref{fig:H4_diffs}). 
Our results regarding the excited states are much more interesting, revealing several important features about the comparison between pCCD and DOCI.
In the first scenario (ground-state HF reference), the striking similarity between pCCD and DOCI energies does not hold for excited states.
While the potential energy curves still share a common behavior, significant energy differences are observed, by as much as 1 Hartree for the quadruply-excited state, $(\sigma_g^*)^2(\sigma_u^*)^2$.
Furthermore, the first [$(\sigma_g)^2(\sigma_g^*)^2$] and the third [$(\sigma_g)^2(\sigma_u^*)^2$] doubly-excited states can be described by two distinct pCCD solutions each.
For the lower-lying one, there are two real solutions close in energy below \SI{1.7}{\bohr}, where they merge into a complex conjugate pair.
The branching point for the higher-lying state is found at \SI{3.3}{\bohr}, but here the complex solutions are found at shorter distances.

When orbitals are variationally optimized for each state at the pCCD level (see bottom panel of Fig.~\ref{fig:H4_both}), the scenario is completely different.
Now, oo-pCCD and DOCI levels of theory provide quite similar energies for all doubly- (and quadruply-) excited states, with visually indistinguishable potential energy curves.
\rev{(Note that the DOCI calculations are also performed with the pCCD optimized orbitals in this case, %.)
such that the two methods are being compared for the same set of orbitals.
Another line of comparison would consider separate orbital optimizations for pCCD and DOCI. 
However, doing so for the latter would be much more computationally demanding,
and we did not explore this scenario here.)}
For all states, oo-pCCD represents a massive improvement with respect to pCCD with HF orbitals.
The average deviations to the corresponding DOCI results drop by one or two orders of magnitude,
while the maximum deviation amounts to 0.02 Hartree, compared with 1 Hartree in the case of HF orbitals (see Fig.~\ref{fig:H4_diffs}).
Not only that, but orbital optimization eliminates the double (and complex) solutions previously discussed for the first doubly-excited state.
While such solutions could arguably be assigned as unphysical, we see that their troubling behavior is as an artifact of the HF reference.
Once this constraint is removed and the orbitals are allowed to relax, only single-valued real solutions appear.
In general, we thus expect that oo-pCCD can locate physical states where non-optimized pCCD fails or gives more dubious results.
 
Our findings can be explained as follows.
The deficiencies of the first set of pCCD calculations can be traced back to the ground-state HF orbitals, which represents a poor reference when excited states are concerned.
Of course, the argument applies for both pCCD and DOCI, but it is more serious for the former, as it lacks higher-order connected excitations (most importantly the connected quadruple excitations).
In particular, these missing excitations account for the largely overestimated pCCD energies associated with the quadruply-excited state.
By optimizing the orbitals at the pCCD level, the reference now provides a qualitatively correct description of each targeted state.
In this sense, more electronic correlation is recovered with the paired double excitations for the optimized reference, in comparison with a ground-state HF reference.
Therefore, higher-order excitations lose importance in oo-pCCD, and the computed energies are very close to the DOCI results.
 
We also found that DOCI is much less sensitive to the orbital choice than pCCD.
The latter energies change by 0.17 Hartree in average, compared with the 0.02 Hartree difference observed in the former.
Again, this can be understood from the lack of higher-order connected excitations in pCCD, which in turn would be partially accounted for by orbital optimization.
On the other hand, all paired excitations are present in DOCI, hence less dependence on the reference.

We notice, however, that oo-pCCD also faced convergence problems when stretching the \ce{H-H} bonds beyond a certain point.
\rev{Either the pCCD equations failed to converge for the targeted excited state or the orbital optimization leads to a solution representing another state.}
This took place for all states, typically at shorter distances as we moved higher in energy.
It could be that a stationary point would only be found if the orbitals had been allowed to become complex, but we did not explore this possibility here.
Instead, this could be a genuine limitation of projected pCCD in describing physical states at geometries with stronger static correlation.
\rev{Finally, symmetry-broken orbitals could perhaps be required at larger bond lengths, but no such solutions were found for the excited states.}

%================================
\subsection{Larger molecules}
\label{subsec:larger}
%================================

The goal of this section is to showcase that $\Delta$oo-pCCD can provide reliable excitation energies for more realistic molecules.
An exhaustive comparison including a larger set of systems and employing different basis sets is beyond the scope of this work.
Here, we have considered five molecules with well-known doubly-excited states, namely \ce{CH+}, \ce{BH}, nitroxyl (\ce{HNO}), nitrosomethane (\ce{H3C-NO}), and formaldehyde (\ce{H2C=O}).
Following previous studies, we employed a bond length of \SI{2.13713}{\bohr} for \ce{CH+}, ~\cite{Olsen_1989,Christiansen_1995b,Kowalski_2001b,Boguslawski_2019}
and of \SI{2.3289}{\bohr} for \ce{BH}. ~\cite{Koch_1995}
For the remaining molecules, the geometries reported in Ref.~\onlinecite{Loos_2019c} were also adopted here.
The frozen-core approximation and the 6-31+G(d) basis set (with spherical gaussian functions) were employed throughout.
For the particular case of \ce{CH+}, we have also considered the basis set presented in Ref.~\onlinecite{Olsen_1989}, and employed in subsequent studies.\cite{Christiansen_1995b,Kowalski_2001b,Boguslawski_2019}

\rev{The set of pCCD optimized orbitals for both ground and doubly-excited states can be found in the {\SupInf}.
These orbitals have been allowed to fully relax and break the molecular point group symmetry,
which took place for the doubly-excited states of nitroxyl, nitrosomethane, and formaldehyde, as well as for the ground state of the latter two.
Doing so can further lower the energy, at the expense of losing the spatial symmetry of the wave function.
We have tried different orbital guesses and damping factors for the orbital optimization, and no other stationary symmetry-broken orbitals were found,
suggesting that the potential problem of multiple solutions would be less serious for the cases studied here.
An inspection of the orbitals confirms that we are describing the $3{}^1\Sigma^+$ and the $2{}^1\Sigma^+$ doubly-excited states of \ce{BH} and \ce{CH+}, respectively, 
in line with previous results. \cite{Olsen_1989,Koch_1995,Christiansen_1995b,Kowalski_2001b,Boguslawski_2019}
Due to the symmetry-broken nature of the orbitals in larger molecules, a precise assignment of the excitation becomes less straightforward, but possible nonetheless.
By analyzing the differences in the orbital densities of ground and excited states, we can infer the
$(n)^2 \rightarrow (\pi^*)^2$ excitations of formaldehyde, nitroxyl, and nitrosomethane, also in consistency with the existing assignments. \cite{Loos_2019c}}

Table~\ref{tab:larger} summarizes the excitation energies for the first excited states with a dominant double excitation character.
While a detailed analysis comprising several levels of theory can be found elsewhere, \cite{Loos_2019c}
here the comparison is focused on our pCCD calculations and EOM-CC based methodologies, 
having the (extrapolated) FCI as the reference result. \cite{Loos_2019c}
For \ce{CH+}, \ce{BH}, and nitroxyl, we managed to find higher-lying solutions of the pCCD amplitude equations, when considering ground-state HF orbitals as reference orbitals.
These solutions were assigned to the targeted doubly-excited states, and are labeled in Table~\ref{tab:larger} as HF-pCCD.
This was achieved by using a very large guess amplitude (typically in the range $\sim 10^2$--$10^3$) for the intended excitation.
At convergence, many other amplitudes attain absolute values larger than 1, yet the one representing the dominant excitation remains the largest.
Higher-lying solutions were also found for nitrosomethane and formaldehyde, yet at significantly higher energies, and it is not clear if they provide a descent representation of the targeted excited state.
Overall, the performance of HF-pCCD is quite erratic, with surprisingly decent results for \ce{CH+} and nitroxyl, but qualitatively wrong ones for \ce{BH}, nitrosomethane, and formaldehyde.
This probably reflects the inadequacy of using ground-state HF orbitals as a reference for describing such states.

\begin{table}[h!]
	\caption{Vertical excitation energies $\Delta E$ (and their corresponding deviation $\Delta\Delta E$ with respect to the FCI reference energies) computed with various methods for the first doubly-excited states of five molecules.
	The 6-31+G(d) basis set was considered throughout, except for the first entry, where the basis set was presented elsewhere. \cite{Olsen_1989}
	Total energies are provided in the {\SupInf}.
	}
\label{tab:larger}
\begin{ruledtabular}
\begin{tabular}{lldd}
molecule    & method  & \tabc{$\Delta E$ (eV)}  & \tabc{$\Delta\Delta E$ (eV)} \\
\hline
\ce{CH+}\fnm[1]
& HF-pCCD &  7.74 & -0.81 \\
& $\Delta$oo-pCCD							&  8.36 & -0.19 \\
& FCI\fnm[2]								&  8.55 & 0 \\
& EOM-CCSDT\fnm[3]							&  8.62 & 0.07 \\
& EOM-CCSDt\fnm[3]							&  8.64 & 0.09\\
& EOM-oo-pCCD-LCCSD\fnm[4]	 				&  8.84 & 0.29 \\
& EOM-pCCD-LCCSD\fnm[4]						&  7.61 & -0.94\\
& CC3\fnm[5]								&  8.78 & 0.23 \\
\hline
\ce{CH+}    
& HF-pCCD        & 7.90    & -0.61 \\
& $\Delta$oo-pCCD & 8.32    & -0.19 \\
& FCI             & 8.51    & 0 \\
& EOM-CCSDTQ      & 8.52    & 0.01 \\
& EOM-CCSDT       & 8.59    & 0.08 \\
& CC3             & 8.75    & 0.24 \\
\hline
\ce{BH}     
& HF-pCCD        & 15.89   & 8.78 \\
& $\Delta$oo-pCCD & 7.35    & 0.24 \\
& FCI             & 7.11    & 0 \\
& EOM-CCSDTQ      & 7.12    & 0.01 \\
& EOM-CCSDT       & 7.15    & 0.04 \\
& CC3             & 7.30    & 0.19 \\
\hline
\ce{HNO}    
& HF-pCCD					& 5.53    & 1.02 \\
& $\Delta$oo-pCCD			& 4.49    & -0.02 \\
& FCI\fnm[6]				& 4.51    & 0 \\
& EOM-CCSDTQ\fnm[6]			& 4.54    & 0.03 \\
& EOM-CCSDT\fnm[6]			& 4.81    & 0.30 \\
& CC3\fnm[6]				& 5.28    & 0.77 \\
\hline
\ce{H3C-NO}  
& $\Delta$oo-pCCD			& 4.66    & -0.20 \\
& FCI\fnm[6]				& 4.86    & 0 \\
& EOM-CCSDT\fnm[6]			& 5.26    & 0.40 \\
& CC3\fnm[6]				& 5.73    & 0.87 \\
\hline
\ce{H2C=O}    
& $\Delta$oo-pCCD			& 11.26    & 0.40 \\
& FCI\fnm[6]				& 10.86    & 0\\
& EOM-CCSDTQ\fnm[6]			& 10.87    & 0.01 \\
& EOM-CCSDT\fnm[6]			& 11.10    & 0.24 \\
& CC3\fnm[6]				& 11.49    & 0.63 \\
\end{tabular}
\end{ruledtabular}
\fnt[1]{Basis set taken from Ref.~\onlinecite{Olsen_1989}.}
\fnt[2]{Results from Ref.~\onlinecite{Olsen_1989}.}
\fnt[3]{Results from Ref.~\onlinecite{Kowalski_2001b}.}
\fnt[4]{Results from Ref.~\onlinecite{Boguslawski_2019}.}
\fnt[5]{Results from Ref.~\onlinecite{Christiansen_1995b}.}
\fnt[6]{Results from Ref.~\onlinecite{Loos_2019c}.}
\end{table}

However, one of the key insights of the present work is that this issue can be largely eliminated by state-specific orbital optimization.
The results labeled as $\Delta$oo-pCCD in Table~\ref{tab:larger} stem from the energy difference between two separate oo-pCCD calculations,
one for the intended doubly-excited state and the other for the ground state.
Not only this method correctly locates all the targeted states, but the excitation energies also compare very favorably with FCI.
For the set of molecules surveyed with the 6-31+G(d) basis set, the mean absolute error (MAE) is \SI{0.21}{\eV}, the root-mean-square error (RMSE) is \SI{0.24}{\eV}, and the mean signed error (MSE) is \SI{0.05}{\eV}.
$\Delta$oo-pCCD is thus considerably more accurate than CC3 (MAE and MSE of \SI{0.54}{\eV}, and RMSE of \SI{0.61}{\eV}),
and comparable to EOM-CCSDT (MAE and MSE of \SI{0.21}{\eV}, and RMSE of \SI{0.25}{\eV}).
Importantly, the excitation energies are not systematically over- or underestimated with respect to the reference values.
In contrast, they tend to be overestimated with EOM-CC methodologies, not only for doubly-excited states but also for singly-excited states.
 
The above observations regarding accuracy and precision of $\Delta$oo-pCCD and EOM-CC methods can be understood 
based on how the reference wave functions are described, and how correlation effects are included in each case.
In EOM-CC, the HF Slater determinant serves as the reference for both ground and doubly-excited states,
thereby creating a bias towards the former.
In addition, excitations are introduced exponentially in the cluster operator, but linearly in the EOM excitation operator, further favoring the ground state.
All in all, electronic correlation is quickly (in the sense of the excitation degree) introduced for the ground state,
while the excited states are still recovering from its poor HF reference.
This is most critical in the case of EOM-CCSD, which largely overestimates (by $\sim$ \SI{1}{\eV} or more) excitation energies of doubly-excited states. \cite{Kowalski_2001b}
Higher orders of excitation are needed in order to obtain more accurate results, yet still they tend to approach the reference values from above.

The situation for $\Delta$oo-pCCD is quite different.
Each state has its own reference, with orbitals variationally optimized for a pCCD wave function.
We notice that this goes beyond a MOM-based pCCD calculation, where the excited state optimized orbitals would have mean-field quality.
Another key distinction is that $\Delta$oo-pCCD provides a unified description of each state, while EOM-based formalisms rely on a formal distinction between them.
Therefore, correlation effects should be accounted for at the same pace, regardless of the targeted state.
This is exactly what we observe, as excitation energies lie either above or below the reference values.
This means that the correlation energy of both ground and excited states are recovered in a balanced way.
Furthermore, paired double excitations already account for important electronic correlation of the doubly-excited states, given that their orbitals have been optimized.
This explains the comparable performances of $\Delta$oo-pCCD and EOM-CCSDT.
In the latter, lower-order excitations would account for orbital relaxation, while actual correlation would only be introduced with the higher-order terms.

Recent studies with state-specific, orbital-optimized density functional theory have also pointed out encouraging results. \cite{Hait_2020,Carter-Fenk_2020,Hait_2021}
RMSEs lie in the range of 0.15--\SI{0.65}{\eV}, depending on the choice of functional.
Bearing in mind the limited set of molecules we explore here, the accuracy of $\Delta$oo-pCCD would be comparable to that of the best performing functional (B97M-V). \cite{Hait_2020,Carter-Fenk_2020}

\rev{While having a dominant doubly-excited character, the states that we have surveyed also present some single excitation contribution, 
however not surpassing 5\% based on CC3 calculations. \cite{Loos_2019c}
As long as this percentage is small, the lack of single excitations is not expected to greatly affect the performance of the $\Delta$oo-pCCD method.
Furthermore, part of this contribution would not correspond to a truly singly-excited character, but rather to orbital relaxation in the excited state, which in turn is accounted for in $\Delta$oo-pCCD.
It remains to be seen how the method performs for doubly-excited states that have a more considerable singly-excited character, as in butadiene and benzene. \cite{Barca_2018a,Barca_2018b,Loos_2019c}}

%%%%%%%%%%%%%%%%%%%%%%%%%%%%%%%%
\section{Conclusion}
\label{sec:ccl}
%%%%%%%%%%%%%%%%%%%%%%%%%%%%%%%%

We have explored excited-state solutions of the pair coupled-cluster doubles (pCCD) method, a version of CCD where the cluster operator is restricted to paired double excitations.
For the helium atom in the 6-31G basis set, we have discussed key aspects regarding the solutions of pCCD and orbital optimization.
In particular, we have shown that the Jacobian matrix of the CC residual equations have to be provided (even if approximately) when aiming for excited states.
Similarly, the orbital rotation Hessian is needed when performing orbital optimization for excited states.

Our first goal was to establish a connection between pCCD and DOCI for excited states.
For that, we have investigated the symmetric dissociation of the linear \ce{H4} molecule with the STO-6G basis set as a function of the distance between the hydrogen atoms.
When the reference is described with ground-state HF orbitals, excited state solutions of pCCD and DOCI no longer match, in contrast to the ground-state case.
Such deviations arise because pCCD struggles more than DOCI in recovering electronic correlation, due to the missing higher-order connected excitations in the former.
However, these higher-order excitations only become important because of the unsuitable starting point provided by the ground-state HF wave function.
By variationally optimizing the orbitals (at the pCCD level) for a targeted doubly-excited state, the reference is significantly improved, and higher-order excitations become much less relevant.
Therefore, when state-specific optimized orbitals were employed, pCCD and DOCI methodologies delivered much closer excited-state energies.

The second goal was to probe the performance of state-specific orbital-optimized pCCD (or $\Delta$oo-pCCD) in describing excited states with strong double excitation character.
We have surveyed a set of five molecules with well-characterized doubly-excited states.
With ground-state HF orbitals, higher roots of pCCD either provides inaccurate excitation energies or fails to locate the targeted state at reasonable energies.
Once again, the problem lies on the reference function, rather than on pCCD itself.
Orbital optimization brings a dramatic improvement, as excitation energies become in much better agreement to the reference (FCI) values,
with a performance superior to CC3 and similar to EOM-CCSDT.

Thus, $\Delta$oo-pCCD might be considered as an alternative and accurate option for targeting doubly-excited states.
Although studies on a larger set of molecules and with other basis sets would be recommended, the current initial results on $\Delta$oo-pCCD are quite promising.
Of course, additional challenges might appear when one considers more chemically challenging situations (larger systems, strong correlation, etc).
\rev{Furthermore, it remains to be seen how the method would perform when one employs state-specific HF orbitals instead of the more expensive optimized orbitals at the pCCD level, as we have done here.
We hope to report further on these aspects in the near future.}

%%%%%%%%%%%%%%%%%%%%%%%%%%%%%%%%
\begin{acknowledgements}
This work was performed using HPC resources from GENCI-TGCC (2020-A0080801738 and 2021-A0100801738) and from CALMIP (Toulouse) under allocation 2021-18005.
This project has received funding from the European Research Council (ERC) under the European Union's Horizon 2020 research and innovation programme (Grant agreement No.~863481).
\end{acknowledgements}
%%%%%%%%%%%%%%%%%%%%%%%%%%%%%%%%

%%%%%%%%%%%%%%%%%%%%%%%%%%%%%%%%%%
\section*{Supporting information available}
%%%%%%%%%%%%%%%%%%%%%%%%%%%%%%%%%%
Raw data associated with Fig.~\ref{fig:H4_both}, total energies associated with Table \ref{tab:larger},
\rev{and state-specific optimized orbitals.}

%%%%%%%%%%%%%%%%%%%%%%%%%%%%%%%%
\bibliography{pCCD}
%%%%%%%%%%%%%%%%%%%%%%%%%%%%%%%%

\end{document}